# A Percolation based M2M Networking Architecture for Data Transmission and Routing

**Jihua Lu, Jianping An, Xiangming Li, Jie Yang and Lei Yang**
School of Information and Electronics, Beijing Institute of Technology
Beijing, 100081 - China
[e-mail: lujihua@ieee.org, an@bit.edu.cn, xiangming_li@ieee.org]
*Corresponding author: Xiangming Li

## Abstract

We propose a percolation based M2M networking architecture and its data transmission method. The proposed network architecture can be server-free and router-free, which allows us to operate routing efficiently with percolations based on six degrees of separation theory in small world network modeling. The data transmission can be divided into two phases: routing and data transmission phase. In the routing phase, probe packets will be transmitted and forwarded in the network thus path selections are performed based on small-world strategy. In the second phase, the information will be encoded, say, with the fountain codes, and transmitted using the paths selected at the first phase. In such a way, an efficient routing and data transmission mechanism can be built, allowing us to construct a low-cost, flexible and ubiquitous network. Such a networking architecture and data transmission can be used in many M2M communications, such as the stub network of internet of things, and deep space networking, and so on.

***Keywords:*** Machine-to-machine, percolation networking, fountain codes, internet of things.

This work was partly presented in IEEE Infocom 2011 Workshop on M2MCN and was supported in part by NSF of China with grants 61002014, 60972017 and 60972018, the Excellent Young Teachers Program of MOE, PRC with grant 2009110120028, the Research Fund for the Doctoral Program of Higher Education with grants 20091101110019, and the Important National Science & Technology Specific Projects with grant 2010ZX03003-004-03 and 2010ZX03002-003-03, and the Beijing Natural Science Fund with grant 4101002.
**DOI:** 10.3837/tiis.0000.00.000

## 1. Introduction

**T**he stub network of internet of things (IoT) may have a lot of sub-netwoks buliding with machine-to-machine (M2M) communication systems, which are usually highly dynamic. That is, networking nodes and links may frequently join to the network and exit from the network. In addition, such networks are often composed using relatively inexpensive nodes that have low power consupmtion, low processing power, and bandwidth. The conventional internet architecture and TCP/IP are not suitable for the the applications of M2M networks because such conventional networks require a lot of servers and routers, and the high overhead in their protocols result in high bandwidth needs which is not reasonable for bandwidth limited communications. Therefore, it is necessary to examine the efficient network architecture and effective data transmission method to achieve better performance with low cost for the M2M communications.

The rest of the paper is organized as follows. In Section 2, we review the previous work in networking, routing and data transmission of M2M communications. Section 3 addresses the M2M network architecture based on six degree separation principle, including the brief introduction of network topology, routing table and network initialization. Section 4 details the efficient percolation routing and data transmission process with their applications. And finally, we conclude our work in Section 6.

## 2. Related Work

To meet the requirements of low-cost, small size, low power applications for M2M equipments (M2MEs), several protocols and physical routing methods are proposed for M2M communications and used in various standards organizations[1-4]. In a M2M network scenario, the devices are usually very small, and are interconnected over wireless communications. Moreover, both the power consumption and processing ability of the nodes are restricted. Moreover, the scale of network may vary dynamically [3-5].

M2M systems can be regarded as an extension of the existing interaction between humans and applications through the new dimension of "things" communication and integration [5][6]. In the M2M systems or networks, machines are clustered together to create a stub M2M network, and are then connected to its infrastructure, i.e., the traditional "Internet of people". For example, researchers predict that by 2014 there will be 1.5 billion wirelessly connected devices that are not mobile phones and do not require any human intervention. As the network scale becomes large, efficient reliable and reliable data transmissions are becoming more challenge. However, it is costly to use the traditional Internet protocol in a M2M scenario. First of all, the M2M network bandwidth is limited and its data packet size is usually small, while the TCP/IP protocol and its ARQ mechanism consume a lot of extra bandwidth. Second, TCP/IP protocol requires a large number of routers and gateways, which increase the complexity of the network structure, leading to high network throughput consumption and maintenance cost [6]. Ad hoc network works without any gateway and special routers [7-10]. However, in Ad hoc, each node must be informed of the complete structure of the entire network to calculate its routing tables. As a result, a lot of control and signal messages should be sent among the network nodes to maintain the routing. Any change in the network structure, as nodes' joining or departure, may generate significant overhead. Furthermore, to manage and operate a large routing table, a node requires rather high level of storing and processing ability.

A M2M network scenario is usually composed of hundreds to thousands of, or even more nodes, and each node can directly communicate with several other nodes. One the other hand, in many M2M networks, the average geodesic distance between any two nodes is relatively short [11]. As indicated by the small world theory, most of the nodes in a very large scale networks may be fairly close to one another. The average distance between pairs of nodes in large empirical networks are often much shorter than in random graphs of the same size.

On the basis of such an observation, we build a small world routing strategy: if a node wants to communicate with a non-conterminous node, it simply hands its adaptive coded data packets through multiple paths selected using percolation method, greedy algorithm or other rules. When the data packets are received by its conterminous nodes that may be neighbors of the destination node, and the conterminous nodes repeat the same procedure until the data is successfully delivered to the destination. To avoid the bandwidth waste, a probe mechanism is used in path selection. Moreover, to prevent the frequent feedback signals in the data transmission, fountain coding is used to encode the source packets. Such a routing and data transmission strategy will be referred to as percolation network scheme.

## 3. Proposed Machine-to-machine Network Architecture Based on "6-degrees" of Distance Rule

A typical M2M network system with small-world properties of M2M network is usually composed of hundreds to thousands of, or even more nodes, and each node can router-freely communicate with several other neighbor nodes. Each node in the network connected to several neighbor nodes randomly. In such a way, a small-world network is formed, in which efficient routing can be operated with percolations based on the six degrees of separation [12].

### 3.1 M2M Network Topology Based on "6-degrees" of Distance Rule

Due to the low-power of M2M devices and the signal-attenuation nature of wireless signals, a M2M node can directly communicate with some nodes nested within networks of neighborhood relations, while it is hard to communicate with the other nodes that are out the scope of radio range [13]. Therefore we take the network structure shown in **Fig. 1** as an example of the M2M network, whose nodes live in a small world where the members are not tightly connected in some part. Each node may have direct connections with one or several other neighbor nodes which we refer this as "close-neighbors". For example, the nodes T, U, O, K and M live in neighborhoods and they are close-neighbors of the node L, which means T, U, O, K and M can send data packets to L directly.

Such a network should have the property of social network properties, in which the relations between R and Q ( also between P and J, and between I and H) may represent the leisure relations. Moreover, T, S, O, K, L, M, N, and U may represent the close relations of a family. In such networks, one node may form ties with leisure or close relations with neighbor nodes nested within them. When the number of close-neighbors of one node is larger than the average close-neighbor number of all the network nodes, we define the node as "Star node", like node G in **Fig. 1**. When one node is not a close-neighbor of another node and if the two nodes want to communicate with each other, they can communicate with each other through relay nodes. Note that, when E wants to communicate with the node N, it may first "guess" that the node L is a close-neighbor of the node E, so it hands its data to the node L. Similarly, the node L "guesses" that the node T may directly connects to the node E, so it forwards the data to the node T. Finally, the node N delivers the data packets to the node E, and vice versa. The one-way or two-way communications are then built between the source and destination nodes based on percolation method and the six degrees of separation principle, and any node can reach everywhere in finite steps of relaying with a very high probability [6][12][13]. In the network, each node acts as relay for the other nodes, it simultaneously transmits its own data packets, thus no dedicated routers are needed. As stated, the routing may be based on "guess", just like the greedy forwarding algorithm which selects the nearest node to the destination node as a relay node within its transmission range, following the rule of the best likelihood [14][15].

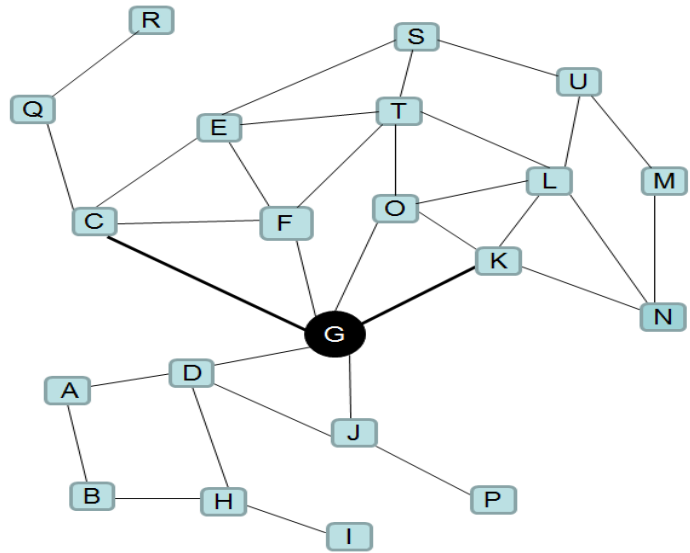

**Fig. 1.** An example of M2M network archtecture with small-world properties

### 3.2 Routing Information Table of M2M network

In **Fig. 1**, each node will hand its data to its close-neighbors which can forward the data to the destination with high probability. Each node may maintain and store a routing information table which contains all or part of the following messages is listed **Table 1**.

**Table 1.** Routing information table of network nodes of a "small-world" M2M network topology

| Field name | Symbol | Descriptions |
| --- | --- | --- |
| Names of close-neighbor nodes | $L_1 L_2 \ldots L_N$ | Name the close-eighbor nodes of the source node as $L$. |
| Number of close-neighbor nodes | $N$ | This field describes the number of nodes available for direct communication without relay. |
| Addresses of close-neighbor | $A_1 A_2 \ldots A_N$ | An address of a node is the identity number of a node. Generally, the name and the address can be the same thus the |

| nodes | | name and the address fields can be merged into one field. |
|---|---|---|
| Bandwidth of the direct links from me to close-neighbors | $R_1 R_2 \ldots R_N$ | Here we define "me" to be the current node. And we suppose that the current node (me) is $C$. We suppose the other nodes are $D_1 D_2 \ldots D_N$. Starting from me, one may have one or more paths connecting to another node with the help of close-neighbors. |
| Measures from me to the node $D_K$ via the close neighbor $L_i$ | $\ldots$ $M_{C\_L_i\_D_K}$ $\ldots$ | Some close-neighbors may take fewer hops while some other close-neighbors may take more hops to reach the destination. The measures from me to a node via a close neighbor reflect the likelihood that a path can be built from me to a destination node through a particular close-neighbor node. To an unknown destination node, one may initialize these values as zero at the first time choosing path for it. |
| Bandwidth from me to the node $D_K$ via the close neighbor $L_i$ | $\ldots$ $R_{C\_L_i\_D_K}$ $\ldots$ | This field enumerate the history data bandwidths from me to another node. For me is unkown the information of destination node, these values can be initialize to a predetermined minimum values. |

### 3.3 Network Initialization Process of M2M network

The network initialization process is as follows.
- A node sends the query signal to its close neighbor nodes. Any node who receives this query signal sends back a reply. The nodes involved in these query and reply signals establish and renew their corresponding information in their information tables, including $N$, $L_i$s, $A_i$ s or $M_{C\_L_i\_D_K}$ s ,etc.
- Set all the measures from me to other nodes via close neighbors to 0 and set all the bandwidth from me to other nodes via close-neighbors to the predetermined minimum values. For those unknown destination nodes to me, reserve adequate memory for the measures and bandwidths.

### 4. Routing and Data Transmission Method Based on Percolation

To provide a reliable route consising of one or multiple paths between any two nodes, link and node outages are very challenging tasks in wireless networks [16][17]. We propose a routing method for the M2M network based on six-degree seperation principle, which is referred to as percolation routing. Suppose that the M2M network has many nodes, and every node has connections with several close-neighbor nodes. The scale of this type of network may change with time, that is, its connection relationship may change with the increasing and disappearing of nodes and link bandwidth may change dynamically with the battery consumption, movement in places of the nodes, and so on.

### 4.1 Introduction of Percolation based Routing in M2M network with Small-world Properties

When a network scale is large and if a node is randomly connected to several nodes in the network, this network can be modeled as a small world network as long as a node is randomly connected to several other nodes. In such a network, when a node wants to send data packets to another node and if the source cannot reach its destination directly, it simply hands the data packets to its close-neighbors which are close to the destination, according to the likelihood /probability rule or the water-pouring principle of power distribution. Thus, several paths are selected via repeating the close-neighbors' chosen procedure until the probe data packets reach the destination node. In such a way, multiple paths between the source and the destination nodes may be built. Notice that some paths will terminate early before they connect to the destination due to the link outages of network jamming and path halt, part of the probing data packets will disappear in the network without reaching the destination node. On the other hand, to avoid frequent ARQ's feedbacks and improve the flexibility of data transmission, the source data may be encoded using fountain codes [18][19][20]. Specifically, fountain code is a class of rateless codes and the data rate varies according to the instant channel state information contrast to the typical fixed-rate code. With fountain codes, the destination is then capable of successfully decoding the received file if a sufficient number of packets is received, regardless of which packets have been exactly received. This process of routing and data transmission is just like the water seeping in a porous material. As little trickle becomes a stream, and stream becomes torrent, multiple parallel data transmission paths may merge to a high-speed route. Due to this reason, following the literature [21], the proposed routing and data transmission based on six-degree separation principle are referred to as percolation based routing and data transmission.

**4.2 Percolation Routing Process in M2M network with Small-world Properties**

It is seen that percolation routing is based on probability thus some paths may reach the destination finally whereas other paths will terminate at other nodes, without reaching the destination. Thus the data transmission is inefficient. To avoid the bandwidth loss, we divide the data delivery procedure into two phases: routing phase and data transmission phase. The former phase corresponds to path selection in which probing packets will be transmitted and relayed in the network. In the later phase, paths are available and data transmission will be performed efficiently. To prevent flooding, the number of hops of a path will be restricted up to $Y$, for example set this value to six. Moreover, to balance the network resource, we limit the number of maximum paths to $X$, for example set this value to five.

Assume that the network initialization is complete, i.e., each node has its initialized routing information table. Then the question of how to discover these paths in an unknown, random wireless network to enable robust multipath routing arises.

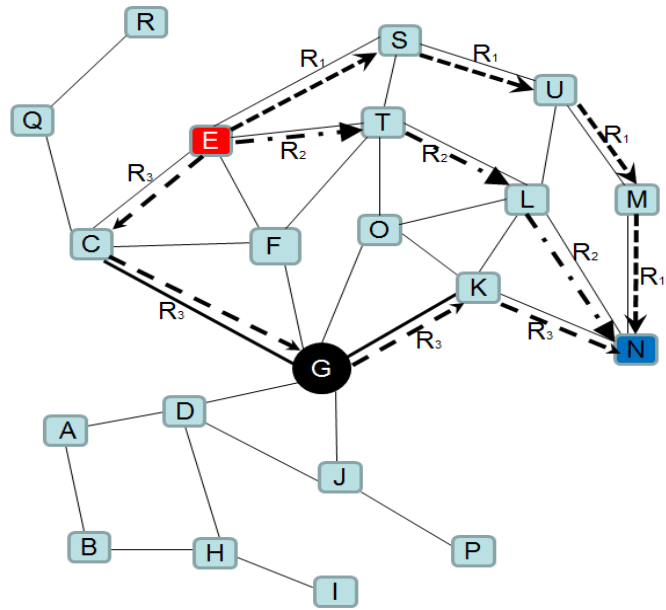

**Fig. 2.**   A routing with multiple paths based on percolation

A stub network structure with its close-neighbor relationships is shown in **Fig. 2**, which includes A, B, C, …U, altogether twenty-one nodes. There are several close-neighbors of each node. For example, node E has four close-neighbors and node A has two close-neighbors. Node I only has one close-neighbor which means that so long as the node I is not the destination node, the probe data packets which have been received by node I would be dropped by node I. Between any two nodes, one or more paths may be built. For example, though direct connections between Node E and N are not available, multiple paths as E-F-G-K-N and E-T-O-L-N can be established between these two nodes.

Then the percolation routing procedure is as follows.

Step 1: The source node sends probe packets to its close-neighbor nodes. Probe packet contains the source address, destination address, and a hop counter. Let us reserve $Y$ positions in the probe for the relay nodes. Set the counter number to zero.

Step 2: A close-neighbor node receives the probe packet. If the node is not the destination node, it will check the hop counter and switch to the following two cases if the hop counter is less than the preset value Y :

Case 1: If there doesn't exist a close-loop close-neighbor node in this selecting path, increase the hop counter by one and the address of the current node is written into the probe packet. Forward the renewed probe packet to a close-neighbor of the maximum measures. Repeat the second step until the destination is reached, then go to the third step.

Case 2: If no any close-neighbor forms a non-closed path, we drop this probe data packet and terminate this routing process.

Step3: The destination node calculates and selects maximum $X$ distinct or parallel paths as the routing. The rule to select paths may be based on the least hops, the maximum throughput, and so on.

Step4: The destination node feeds back the acknowledgement packets composed of the selected path information to the transmitter node through the selected paths. When the feedback information passes through the intermediate nodes, each intermediate node records the routing information and renews its routing information table by increasing each measure from me to any involved node on the path via close neighbors with a unit value.

Thus, the routing procedure is complete. As seen in **Fig. 2**, a route consisting of three paths $R_1$, $R_2$, and $R_3$ is selected. Now it is ready to transmit the data packets.

## 4.4 Percolation based Data Transmission with Fountain Codes

Then the data file will be transmitted in the following procedure.

Step 1: The source node encodes the data file to be transmitted using fountain codes, as LT codes and Raptor codes. Assign the loads to the paths proportional to their throughput capacity. Feed the encoded packets to the selected paths according to their load capacity.

Step 2: An intermediate node receives the data packets. If the node isn't the destination node, the data packets will be forwarded to the next node on the path, until the destination is reached, otherwise, continue to the third step.

Step 3: The destination node will assemble the data packets received and decode them. An ACK signal will be sends back to the transmitter node as soon as the data file is successfully recovered.

Step 4: The source node receives the ACK from the destination node and go to step 1 to encode and transmit the next data frame.

Suppose a route consisting multiple percolation paths between the transceiver nodes, like the established multiple paths in **Fig. 2**. Each percolation path sends proper number of coded packets. The packets organization and their encoding for percolation routing and data transmission is shown in **Fig. 3**. First, the data of length $K \cdot M$ are portioned into $K$ input packets, i.e., the length of each input packet is $M$. The $K$ source packets $(a_0^{(0)}, a_1^{(0)}, ..., a_{M-1}^{(0)}), (a_0^{(1)}, a_1^{(1)}, ..., a_{M-1}^{(1)})$, ... , $(a_0^{(K-1)}, a_1^{(K-1)}, ..., a_{M-1}^{(K-1)})$ are reorganized in the column order as $(a_0^{(0)}, a_0^{(1)}, ..., a_0^{(K-1)})$, $(a_1^{(0)}, a_1^{(1)}, ..., a_1^{(K-1)})$, ... $(a_{M-1}^{(0)}, a_{M-1}^{(1)}, ..., a_{M-1}^{(K-1)})$. Then, each reorganized packet $(a_i^{(0)}, a_i^{(1)}, ..., a_i^{(K-1)})$ is fountain encoded into a semi-infinite stream $(A_i^{(0)}, A_i^{(1)}, ..., A_i^{(N-1)}, ...)$, where $i = 0, 1, ..., M-1$. Reading the fountain-coded packets in the column order gives the output of the semi-infinite packets stream as $(A_0^{(0)}, A_1^{(0)}, ..., A_{M-1}^{(0)})$, $(A_0^{(1)}, A_1^{(1)}, ..., A_{M-1}^{(1)})$, ..., $(A_0^{(N-1)}, A_1^{(N-1)}, ..., A_{M-1}^{(N-1)})$, ..... The fountain codes used in encoding can be LT code, or Raptor codes [19][20].

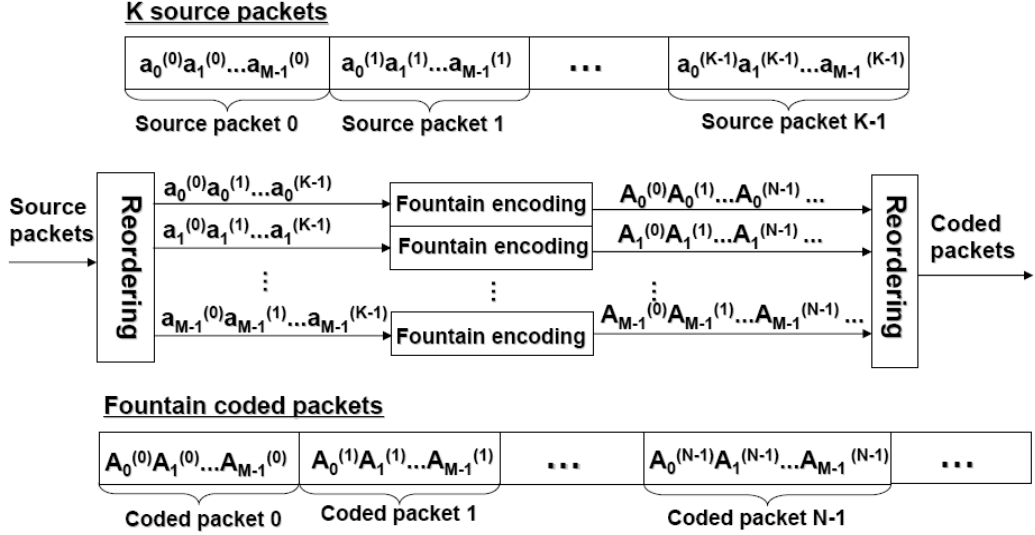

**Fig. 3.**  Packets organization and their encoding scheme

When the destination node receives enough coding packets to decode the K source packets, it will send "stop sending" signal back to the transmitted source node. Then the transmitter stops sending the coded packets to receiver/destination node. If no feedback channels are available for termination signals, fountain codes can work in forward error correction (FEC) mode which provides a reliable data transmission.

### 4.5 Applications of the Proposed Percolation method

The proposed percolation based routing and data transmission can be used in many M2M communications, including but not limited to the following:

- The stub network of IoT, such as internet of cars, internet of goods for supermarket chains, and internet of medical instruments. Such networks are highly dynamical and are usually composed of a lot of nodes, where each node can directly communicate with several other nodes. Percolation based routing and data transmission can work without any dedicated routers and servers, providing a low-cost, flexible and high efficient networking and data transmission deployment.
- Deep space networks for communications with relaying nodes of man-made satellites. In such networks, a relaying node of satellite may join to or exit the network frequently due to the revolution and rotation of planets, and the satellite's running. The percolation based routing provides multiple paths. Since the fountain encoded data streams are transmitted through multiple paths, even if part of the links become unavailable, reliable data transmission can be guaranteed between the deep space ship and the earth.  An example network for the earth-moon communications is illustrated in **Fig. 4**. In the network, several man-made earth relaying satellites and moon relaying satellites form a relaying network. A route of multiple paths can be built from the earth to the moon through the relaying network. It is seen that an earth base station can communicate with a robot on the moon reliably via multiple paths using the proposed percolation based routing and data transmission.
- Anti-jamming communications. A typical M2M network may consist of a lot of distributed network nodes. Part of the nodes may be exposed to interference and jamming. The nodes experiencing heavy interference may become unavailable, which may lead to path break or packet loss. Since the fountain codes are combined with multiple-path routing, reliable communications can be guaranteed. Therefore, the use of percolation routing and data transmission in such a network is helpful to increase the anti-jamming performance.

- Distributed data storage and safety. In a M2M network, one node can encode its data using the fountain codes and then cut the encoded data into slices. The data slices will be distributed and stored into many neighbor nodes using the percolation method. This storage will be robust and safety. For example, a data file of K source packets can be first fountain encoded. For safety, the original node will store $S_1$ encoded packets into itself memory and distribute $S_2$ encoded packets to its neighbors for storage. If $S_1+S_2>K$ and $S_2<K$, the source can recover the source file while any other nodes are not likely to recover the source file.

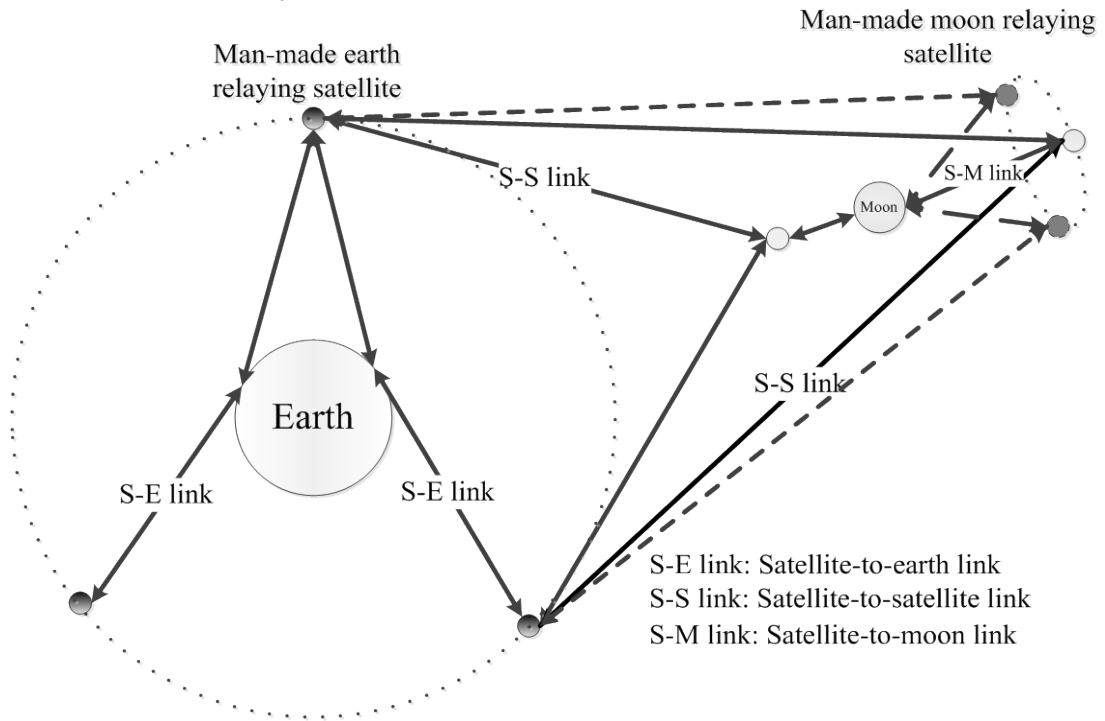

**Fig. 4.** Percolation based routing and data transmission in Earth-Moon man-made satellite relaying system

## 5. Conclusions

We proposed a percolation based routing and data transmission method for the M2M network using the six degrees of separation principle. Under the network structure with "6-degree" distance properties, the proposed scheme consists of two phases: routing phase based on percolation and data transmission phase based on fountain codes. In the routing phase, probes packets are transmitted and flowed in the network, multiple paths are built which form a route. After that, the data file will be fountain encoded and the transmitted data rate is adaptively varying according to the real time channel state information. It is seen that the proposed network routing method either works when a node has the full network architecture or it does when a node contains only local network structure. Since the storage and processing ability of a node in a M2M network is limited, it works with limited information of local close-neighbors. The proposed method has the advantages of efficiency, self-management, self-maintaining and low cost in network deployment and distributed data storage.